# Measuring Triple-Helix Synergy in the Russian Innovation Systems at Regional, Provincial, and National Levels

*Journal of the Association for Information Science and Technology JASIST* (in press)

Loet Leydesdorff,[a] Evgeniy Perevodchikov,[b] and Alexander Uvarov[c]

[a] corresponding author; Amsterdam School of Communication Research (ASCoR), University of Amsterdam, Kloveniersburgwal 48, 1012 CX Amsterdam, The Netherlands; tel.: +31-20-693 0565; fax: +31- 842 239 111; email: loet@leydesdorff.net ;

[b] Tomsk State University of Control Systems and Radioelectronics (TUSUR), Institute for Innovations, 40 Lenina Prospect, Tomsk, 634050, Russia; tel.: +7 (3822) 510536; fax: +7 (3822) 510536; evp@2i.tusur.ru

[c] Tomsk State University of Control Systems and Radioelectronics (TUSUR), Institute for Innovations, 40 Lenina Prospect, Tomsk, 634050, Russia; tel.: +79138290661; fax: +7 (3822) 510536; email: au@tusur.ru

**Abstract**
We measure synergy for the Russian national, provincial, and regional innovation systems as reduction of uncertainty using mutual information among the three distributions of firm sizes, technological knowledge-bases of firms, and geographical locations. Half a million data at firm level in 2011 were obtained from the *Orbis*™ database of Bureau Van Dijk. The firm level data were aggregated at the levels of eight Federal Districts, the regional level of 83 Federal Subjects, and the single level of the Russian Federation. Not surprisingly, the knowledge base of the economy is concentrated in the Moscow region (22.8%); St. Petersburg follows with 4.0%. Only 0.4% of the firms are classified as high-tech, and 2.7% as medium-tech manufacturing (NACE, Rev. 2). Except in Moscow itself, high-tech manufacturing does not add synergy to any other unit at any of the various levels of geographical granularity; instead it disturbs regional coordination even in the region surrounding Moscow ("Moscow Region"). In the case of medium-tech manufacturing, there is also synergy in St. Petersburg. Knowledge-intensive services (KIS; including laboratories) contribute 12.8% to the economy in terms of establishments and contribute to the synergy in all Federal Districts (except the North-Caucasian Federal District), but only in 30 of the 83 Federal Subjects. The synergy in KIS is concentrated in centers of administration. Unlike Western European countries, the knowledge-intensive services (which are often state-affiliated) thus provide backbone to an emerging knowledge-based economy at the level of Federal Districts, but the economy is otherwise not knowledge-based (except for the Moscow region).

1. **Introduction**

Following the demise of the Soviet Union, the Russian state apparatuses and the Russian innovation system were disorganized during the 1990s. More recently, innovation policies and the construction of innovation systems are back on the agenda: Can one use the wealth from oil and gas revenues to convert the Russian economy from a resource-dependent into a knowledge-based one? How should one stimulate innovation in such a vast country? Should this be left to the regions or be coordinated more nationally?

In the last decade, the Russian government has proposed a series of initiatives to stimulate the transition of the country from a resource-based to a knowledge-based economy. Three stages of federal involvement in the development of an innovation infrastructure can be distinguished: (1) during the years 2005-2008, the federal government established science and technology parks across Russia; (2) during 2009-2011, legislature was passed to facilitate the development of innovations; and (3) in the years 2011 and 2012, the government provided federal grants to establish entrepreneurial universities and regional innovation clusters (Perevodchikov and Uvarov, 2012).

The government actively encourages the development of science and technology parks, technology transfer offices, innovation centers, spin-off programs, etc. Technology incubation centers were set up within local universities, and their entrepreneurial capacity and ability to generate, manage, and promote start-ups were encouraged. Skolkovo, a large innovation and education center to be built in Moscow in collaboration with the Massachusetts Institute of Technology, is the most recent government initiative. In the years 2012-2014, the government plans to invest almost $3 billion in this development (Pakhomova, 2013).

In 2009-2010, amendments were passed to Federal Law 217 (Federal Law 217-FZ, 2009) and Federal Decrees 218-220 (Federal Decrees 218-220, 2010) were issued in order to create opportunities for effective collaborations between universities, business, and government(s). Federal Law 217 is similar to the Bayh-Dole Act in the USA (e.g., Mowery & Sampat, 2004) and has the objective to create a collaborative environment between universities and companies, by encouraging involvement of scientific and technological institutions in the innovation process. Federal Decree 219 awarded up to $267 million during 2010-2012 to support the development of infrastructures for innovation in the higher-education sector; Federal Decree 220 awarded $400 million in 2010-2012 to support leading scientists.

The recent government priorities are directed towards facilitating trilateral collaborations in university-industry-government relations as a part of Russia's strategic economic development. A related issue remains the allocation of national government funding to stimulate innovations at the regional level. The Russian economy can nowadays be considered as a combination of free-market activities with a variety of government interventions. Abundant in natural resources, Russia remains highly dependent on the petroleum sector; the role of the state in stimulating the transition to a knowledge-based economy remains crucial given this condition.

## 2. Synergy in three dimensions

To what extent—in which regions and sectors—can a synergy in innovation systems be indicated in Russia? The main objectives of this study are to analyze the knowledge-based economy of Russia and measure the quality of the Russian (regional) innovation systems using the indicator of synergy based on entropy statistics (Jakulin & Bratko, 2004; see also Guo, 2010; Chanda *et al.*, 2007; Lewontin, 2000: 10-12; Yeung, 2008). A number of European studies have estimated the reduction of uncertainty at the systems level using this Triple-Helix indicator of the (latent) synergy in the knowledge base of an economy, including the Netherlands (Leydesdorff, Dolfsma, & Van der Panne, 2006), Sweden (Leydesdorff, & Strand, 2013), Germany (Leydesdorff and Fritsch, 2006), Hungary (Lengyel & Leydesdorff, 2011), and Norway (Strand & Leydesdorff, 2013). Studies about China (Leydesdorff & Zhou, in press) and Italy (Cucco & Leydesdorff, in preparation) are also forthcoming.

The Triple Helix indicator of synergy is more abstract than the Triple Helix model of university-industry-government relations; the specification of institutional relations can be considered as a first step in the operationalization of an eco-system (Etzkowitz, 2001 and 2007; Storper, 1997). However, synergy among three functions such as novelty production, wealth generation, and normative control in innovation systems is based on correlations—instead of relations—among distributions in three (or more) dimensions. University-industry-government relations, for example, can be considered as an eco-system of bi-and trilateral relations in which these functions can operate on the basis of the correlations that are consequently shaped at the network level.

In this study (and following the above-mentioned ones) we operationalize the function of governance in terms of the distribution of geographical addresses—regions, provinces, nation(s); the economic dynamic of wealth generation in terms of the distribution of firm sizes—small and medium sized enterprises (SMEs) versus large corporations; and the distribution of technological capacities is operationalized in terms of the NACE codes of the OECD.[1]

Firms are assumed as the units of analysis: each firm has an address and thus falls under the jurisdiction of a local, regional, and national government; each firm is attributed NACE codes and has a size in terms of numbers of employees. These three distributions are analytically independent, but they relate in terms of co-variations. For example, firms can be clustered in regions or technological sectors of the economy. The research question is whether the distributions at various levels of administration and/or in sectors of the economy (e.g., knowledge-intensive services) are synergetic in terms of the fits among the distributions?

The mutual information in three (or more) dimensions follows from Shannon's (1948) information theory (e.g., McGill, 1964), but cannot be given an interpretation within this theory because its value can potentially be negative (Krippendorff, 2009a; Yeung, 2008, pp. 59f.). Instead of generating uncertainty (that is, Shannon-type information), redundancy is then generated in next-order loops among the codes in different dimensions. In other words, redundancy is generated because the same instances of co-variation (relation) are provided with different meanings and may therefore be counted more than once. Leydesdorff & Ivanova (2014) argue that one measures not "mutual information," but "mutual redundancy" in three (or more) dimensions.

---

[1] NACE is an abbreviation of *Nomenclature générale des Activités économiques dans les Communautés Européennes*. The NACE code can be translated into the International Standard Industrial Classificiation (ISIC).

In the Triple Helix framework, Etzkowitz & Leydesdorff (2000) suggested that an overlay of shared meanings provided to the relational events may reduce uncertainty depending on the configuration in the total set of relations. The overlay is the result of a configuration of relations, but may begin to feedback on the further shaping of relations. Because of this (partial) sharing of meanings, a niche can be formed in which the prevailing uncertainty is reduced locally. The localization of these niches can be in terms of geography, the technological capacities or the economic dynamics. Empirically, there is always a trade-off between forward uncertainty (variation) generation in relations, and the possible reduction of uncertainty in terms of feedback by the overlay of interacting meanings that operates selectively (Ulanowicz, 2009).

## 3. Operationalization in terms of Shannon's formulas

For each of the three dimensions, our data contains a proxy: geography is indicated with the postal address; economic weight with the size of the firm in terms of number of employees (e.g., SMEs versus large corporations); and technological capacity is indicated in terms of the NACE code (Rev. 2) developed for this purpose by the Organization of Economic Co-operation and Development (OECD) in Paris.

According to Shannon (1948), probabilistic entropy provides a measure of the expected uncertainty in a probability distribution $p_x$ of a variable $x$ as follows: $H_x = -\sum_x p_x \log_2 p_x$. (If two is used as the base of the logarithm, uncertainty is expressed in bits of information.) Analogously, the uncertainty in two dimensions based on the joint probability distribution $p_{xy}$ of two variables $x$ and $y$, is $H_{xy} = -\sum_x \sum_y p_{xy} \log_2 p_{xy}$. However, if there is interaction between the two variables (e.g., the locations and the sizes of firms), this uncertainty is reduced with the mutual information or transmission ($T_{xy}$) as follows: $T_{xy} = (H_x + H_y) - H_{xy}$. (If the distributions are completely independent $H_{xy} = H_x + H_y$ and $T_{xy} = 0$.)

Likewise, the measure of uncertainty in three dimensions based on the joint probability distribution $p_{xyz}$ of three variables $x$, $y$, and $z$ is $H_{xyz} = -\sum_x \sum_y \sum_z p_{xyz} \log_2 p_{xyz}$. It can be shown (e.g., Abramson, 1963: 131 ff.) that the mutual information or transmission is specified as follows:

$$T_{xyz} = H_x + H_y + H_z - H_{xy} - H_{xz} - H_{yz} + H_{xyz} \tag{1}$$

Depending on the relative weights of the terms in Eq. 1, the resulting value of $T_{xyz}$ can be positive or negative (Yeung, 2008, pp. 59 ff.), whereas $T_{xy}$ in two dimensions is always positive. Krippendorff (2009a) showed that Shannon-type interaction information in three dimensions can be approached differently, namely as the (positive) uncertainty added to the sum of the two-way interactions.[2] However, the *signed* information measure $T_{xyz}$ can be considered as the result of next-order loops that may entail positive or negative redundancies (Krippendorff, 2009b, at p. 676). Adding redundancy (other options) to a system adds to the maximum entropy and can thus reduce the relative uncertainty. The signing remains consistent with Shannon's (1948) information theory only if this reduction of uncertainty prevailing at the systems level is subtracted as negative information (Leydesdorff, 2010; Leydesdorff & Ivanova, 2014).

---

[2] Krippendorff (1980; 2009a) suggested to write this Shannon-type information as $I_{ABC \to AB:AC:BC}$.

|      | x | y | z |
|------|---|---|---|
| Nr 1 | 0 | 1 | 0 |
| Nr 2 | 1 | 0 | 1 |
| Nr 3 | 1 | 1 | 0 |
| Nr 4 | 0 | 1 | 1 |

**Table 1**: Example of mutual information in three dimensions

Table 1 provides an example of three binary variables ($x$, $y$, and $z$) attributed to four cases. The uncertainty in $x$ ($H_x$) is precisely one bit of information because the variable is attributed 0 in two cases and 1 in the other two cases: $p(0) = p(1) = 0.5 \rightarrow H_x = -½ \log_2 (½) - ½ \log_2 (½) = 1$ bit. The four combinations of $x$, $y$, and $z$, however, are all unique and hence $H_{xyz} = 4 * [-¼ \log_2 (¼)] = 2$. Using Eq. 1, it follows that in this case $T_{xyz} = 1.00 + 0.81 + 1.00 - 1.50 - 2.00 - 1.00 + 2.00 = -0.19$ bits. Leydesdorff, Park & Lengyel (in press) provides other examples; Sun & Negishi (2010) discuss the potentially negative value of $T_{xyz}$ in relation to partial correlations. A routine for computing mutual information in three or four dimensions is available online at http://www.leydesdorff.net/software/th4 .

One advantage of information theory is that all values are based on summations and can be fully decomposed into the contributing terms. As in the decomposition of probabilistic entropy (Theil, 1972: 20f.), the mutual information in three dimensions can be decomposed into $G$ groups as follows:

$$T = T_0 + \sum_G \frac{n_G}{N} T_G \qquad (2)$$

When one decomposes, for example, a country in terms of its regions (or provinces), $T_0$ is between-region uncertainty or a measure of the dividedness among the regions; $T_G$ is the uncertainty at the geographical scale $G$; $n_G$ the number of firms at this geographical scale $G$; and $N$ the total number of firms in the dataset. The values for $T$ and $T_G$ can be calculated from the respective distributions (using Eq. 1), if the sum values of $N$ and $n_G$ are known. The normalized values of the contributions of regions to the national synergy ($\Delta T = \frac{n_G}{N} * T_G$) and the between-group synergy ($T_0$) can then be derived. $T_0$ is equal to the difference between the $T$-value for the whole set minus the sum of the subsets.

Note that $T_0$ can have positive or negative signs, and can also be expressed as a percentage contribution to the total synergy for a system of reference (e.g., at the national level). A negative value of $T_0$ indicates that the uncertainty at the next geographically aggregated level is reduced more than the sum of the parts, whereas a positive value indicates that the next level of integration does not add synergy to the system. Thus, one can test, for example, whether the national level adds to the systems integration more than the sum of the regional units. The relative contributions at each level can be specified after proper normalization for the number of firms.

## 4. Data

Almost a million records with a Russian address were harvested at the firm level from the *Orbis* database (available at https://orbis.bvdinfo.com) on January 20, 2013 (Table 2). *Orbis*™ is a database maintained by Bureau Van Dijk (BvD) and consists of company-level data for more than 100 million firms (including banks) collected worldwide for commercial purposes. Regardless of its numerous drawbacks and data coverage issues (Ribeiro, Menghinello, and Backere, 2010) —this is not complete governmental statistics—we used this data given a lack of alternatives. In a study of the Italian innovation system (Cucco & Leydesdorff, in preparation), an almost perfect correlation was found (Pearson $r = 0.98$; Spearman's $\rho > 0.99$) between the distributions of the synergy values based on 462,316 valid observations using *Orbis* data versus 4,480,473 firms registered by Statistics Italy in 2007. This gives us some confidence in the representativeness of the *Orbis* data and its usefulness for this purpose.

| Year | N of firms |
|------|-----------:|
| 2008 | 111,943 |
| 2009 | 116,408 |
| 2010 | 170,911 |
| 2011 | 593,987 |

**Table 2**: Distribution of Russian firms with valid data in the *Orbis* database by year.

By far the most complete data at the date of the download was for 2011 (Table 2). For this year, 613,018 records were retrieved, with 593,987 (96.9%) records containing valid information in the three relevant dimensions (address, size, and NACE code). Despite the limitations of this sample, we are able to address our main research question about the synergy generated in the Russian economy in terms of combinations among geography, technology, and organization. Three variables attributed to each firm will be used as proxies for each specific combination in the geographical, technological, and organizational dimensions: (1) the zip code as indicator of the firm's geographical location; (2) size, measured by the number of employees, as a proxy of the economic dynamics—one can distinguish among small, medium (e.g., SMEs), and large firms—and (3) the NACE code (Rev. 2) of the OECD as a measure of the main technology.

**Table 3**. Distribution of Firms by Geographical Region and Federal District.

| Geographical region | Federal District | Number of Firms | % |
|---|---|---|---|
| European part | Central | 227,476 | 38.3 |
| | Volga | 105,699 | 17.8 |
| | Northwestern | 54,671 | 9.2 |
| | Southern | 48,843 | 8.2 |
| | North Caucasian | 10,876 | 1.8 |
| Siberia | Siberian | 80,437 | 13.5 |
| | Ural | 42,090 | 7.1 |
| | Far Eastern | 23,895 | 4.0 |
| ***Total*** | | 593,987 | 100.0 |

A firm's geographic location is captured by the four-digit zip code. This code allows us to aggregate the data into three regional classification levels: national (the Russian Federation), district (8 Federal Districts), and regional (83 Federal Subjects or states) levels (Table 3 and Figure 1). One can also distinguish between the European and Siberian parts of the Russian Federation; but this division is not used by the administration.

More than 75% of the firms (447,565) are located in the European part of Russia as can be expected given the low population rates in the Siberian regions (Table 4). One third of the firms (179,009 or 31%) is located in Moscow, Saint Petersburg, and the Moscow region ("Moscow Region"), reflecting the highly centralized and clustered economic structure of Russia.

Novosibirsk Region has with 19,225 (3%) the largest concentration of firms east of the Ural mountains. The normalized Gini coefficient of the distribution of firms over the 8 Federal Districts is 0.49, and 0.59 over the 82 Federal Subjects.

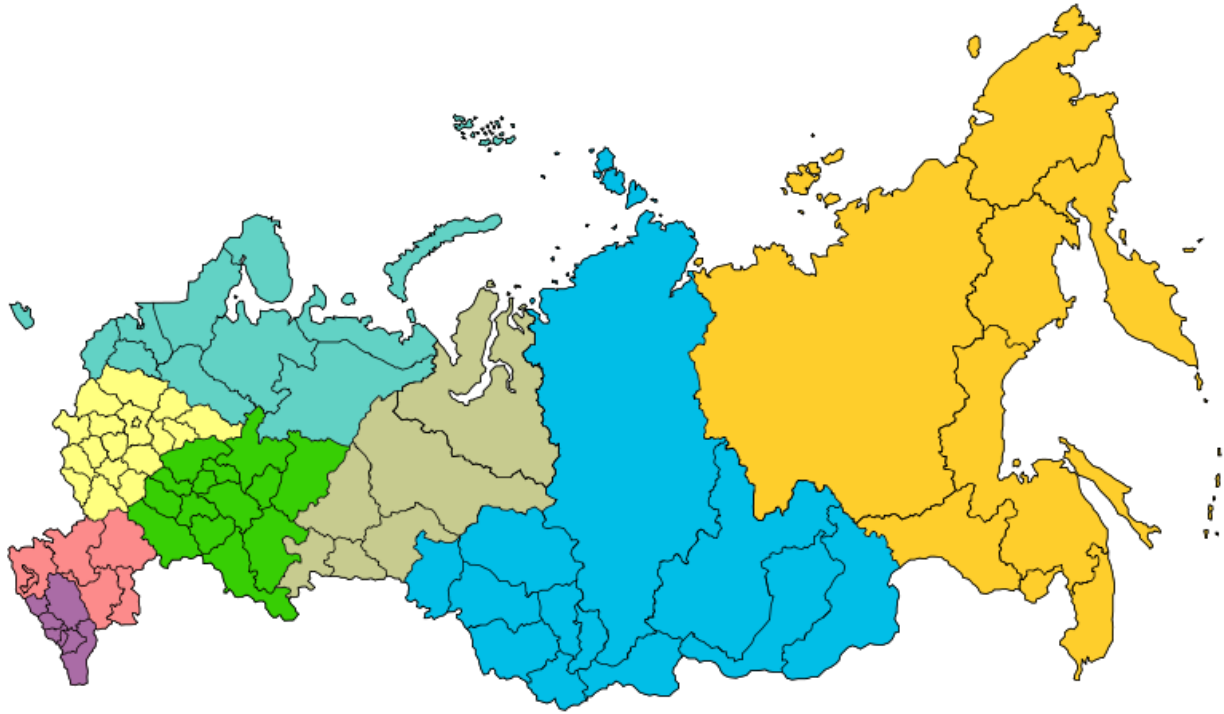

**Figure 1**: Administrative map of the Russian Federation: 8 Federal Districts and 83 Federal Subjects.

| Federal District | Area (km²) | Population (2010 Russian Census) | Federal subjects |
|---|---|---|---|
| Central Federal District (Europe) | 652,800 | 38,438,600 | 18 |
| Southern Federal District (Europe) | 418,500 | 13,856,700 | 6 |
| Northwestern Federal District (Europe) | 1,677,900 | 13,583,800 | 11 |
| Far Eastern Federal District (Siberia) | 6,215,900 | 6,291,900 | 9 |
| Siberian Federal District (Siberia) | 5,114,800 | 19,254,300 | 12 |
| Ural Federal District (Siberia) | 1,788,900 | 12,082,700 | 6 |
| Volga Federal District (Europe) | 1,038,000 | 29,900,400 | 14 |
| North Caucasian Federal District (Europe) | 170,700 | 9,496,800 | 7 |

Source: President of the Russian Federation (2000).

**Table 4:** Administrative organization of the Russian Federation: 8 Federal Districts and 83 Federal Subjects.

The number of employees of a firm can be considered as a proxy of its size and industrial organization (Blau and Schoenherr, 1971). Almost half of the sample consists of medium-sized firms with between 10 and 50 employees (Table 5). According to the European Commission's (2003; 2011) classification of firms by number of employees, micro-entities have less than 10 employees, small-sized firms have less than 50 employees, and medium-sized less than 250.

Companies without employees (including sole ownership firms),[3] represent only a relatively small share of the sample (20%), which is similar to the previously reported Dutch data (19.7%), but well below that in the Norwegian (61%) or Swedish (more than 70%) case studies, respectively. However, we are uncertain about the inclusion of firms with no employees in the *Orbis* dataset (Cucco & Leydesdorff, in preparation; Ribeiro *et al.*, 2010). We followed the discretization in terms of size classes provided by Leydesdorff *et al.* (2006, at p. 186), except for the two largest classes which were chosen so that they are approximately of the same magnitude and in accordance with the definitions of the European Commission (2003; 2011). Classification of the numbers in discrete classes is needed given our information-theoretical framework.

---

[3] It is assumed that zero number of employees implies the sole ownership firm, even if it can be a missing value (Ribeiro *et al.*, 2010).

**Table 5.** Size Classes of Firms in Terms of Numbers of Employees.[4]

| EMPLOYEES | Number of Firms | Percentage |
|---|---|---|
| 0 | 121,553 | 20% |
| 1-4 | 74,892 | 13% |
| 5-9 | 19,978 | 3% |
| 10-19 | 115,853 | 19% |
| 20-49 | 165,246 | 28% |
| 50-99 | 44,822 | 8% |
| 100-249 | 28,502 | 5% |
| >249 | 23,343 | 4% |
| **Total** | **594,189** | **100%** |

The data contains the NACE codes at the four-digit level for each firm, but we use only the first two digits. This provides us with 35 classes in use in this data. The NACE codes can be considered as a proxy of a firm's technology. Using the NACE classification of Table 6, 76,078 (12.8%) of the Russian firms can be considered as knowledge-intensive services (KIS); 2,564 (0.4%) are classified as high-tech manufacturing firms; and 15,860 (2.7%) as medium-tech manufacturing.

**Table 6**: NACE classifications (Rev. 2) of high- and medium-tech manufacturing, and knowledge-intensive services; according to Eurostat/OECD.

| | |
|---|---|
| *High-tech Manufacturing* | *Knowledge-intensive Sectors (KIS)* |
| **21** Manufacture of basic pharmaceutical products and pharmaceutical preparations | **50** Water transport, |
| **26** Manufacture of computer, electronic and optical products | **51** Air transport |
| **30.3** Manufacture of air and spacecraft and related machinery | **58** Publishing activities, |
| | **59** Motion picture, video and television programme production, sound recording and music publishing activities, |
| | **60** Programming and broadcasting activities, |
| *Medium-high-tech Manufacturing* | **61** Telecommunications, |
| **20** Manufacture of chemicals and chemical products | **62** Computer programming, consultancy and related activities, |
| **25.4** Manufacture of weapons and ammunition | |
| **27** Manufacture of electrical equipment, | **63** Information service activities |
| **28** Manufacture of machinery and equipment n.e.c., | **64 to 66** Financial and insurance activities |

---
[4] Of these 594,189 firms, 593,987 (> 99.9) were included into the analysis. Some firms were removed because of missing or incomplete data in one of the other two dimensions.

| | |
|---|---|
| **29** Manufacture of motor vehicles, trailers and semi-trailers,<br>**30** Manufacture of other transport equipment **excluding 30.1** Building of ships and boats, and **excluding 30.3** Manufacture of air and spacecraft and related machinery<br>**32.5** Manufacture of medical and dental instruments and supplies | **69** Legal and accounting activities,<br>**70** Activities of head offices; management consultancy activities,<br>**71** Architectural and engineering activities; technical testing and analysis,<br>**72** Scientific research and development,<br>**73** Advertising and market research,<br>**74** Other professional, scientific and technical activities,<br>**75** Veterinary activities<br>**78** Employment activities<br>**80** Security and investigation activities<br>**84** Public administration and defence, compulsory social security<br>**85** Education<br>**86 to 88** Human health and social work activities,<br>**90 to 93** Arts, entertainment and recreation<br><br>Of these sectors, **59 to 63, and 72** are considered *high-tech services*. |

*Sources:* Eurostat (2014; cf. Laafia, 2002).

The Russian sample has a relatively low share of firms with knowledge-intensive services when compared with the data reported in the Norwegian, Dutch, or German case studies (44%, 51%, and 33%, respectively). High-tech manufacturing is scarce and highly centralized in Moscow itself (Gini = 0.73). Medium-tech manufacturing (2.7%) is more evenly distributed across the regions (Gini = 0.64), whereas the knowledge-intensive services (12.8%) are concentrated in Moscow and its metropolitan environment (Gini = 0.71).

A shape file for mapping the administrative organization of Russia was retrieved from http://www.diva-gis.org/datadown for the geographic mapping using SPSS v20. The database of this file distinguishes the two administrative layers within Russia: eight Federal Districts and 83 Federal Subjects (states, provinces, and republics). Our data does not contain one of the latter subjects (Kostroma Region), and thus we use 82 lowest-level regional units.

Analogously to the previous studies, we also explore whether regional differences in the configurations are determined by high- and medium-tech manufacturing. Research questions are consistent with previous case studies: (1) is the in-between group reduction of uncertainty at the regional (that is, Federal Subject) level larger than at the level of Federal Districts? and (2) is medium-tech manufacturing associated with synergy more than high-tech firms? On the basis of the previous studies, knowledge-intensive services can be expected to uncouple from regional economies because they are less geographically constrained in terms of buildings, etc. However, this may be different for Russia because research laboratories are also classified as knowledge-intensive services.

## 5. Results

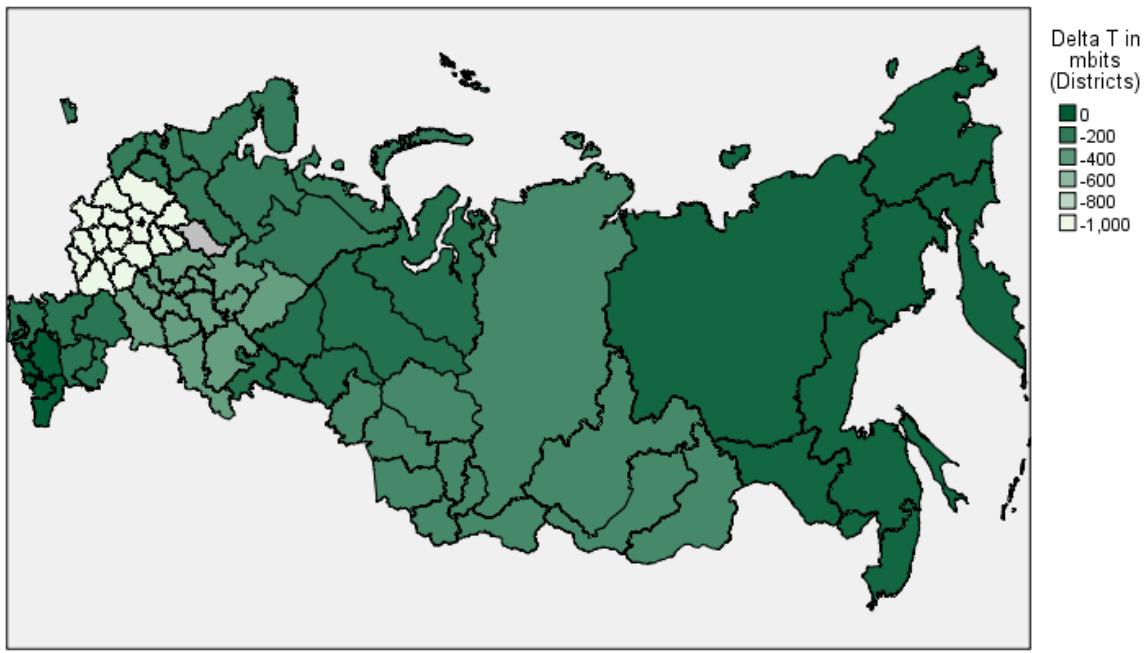

**Figure 2.** Synergies in the knowledge-based economy of Russia ($\Delta T$ in mbits) at the level of the eight Federal Districts (2011; $N = 593{,}987$).

Figure 2 shows that synergy at the level of the eight Federal Districts is found in the European part of Russia more than in the Siberian part. This is not a surprise given the larger density of the firms and volume of government subsidies in the European part of Russia. Within the framework of the Triple-Helix theory, this part of the country has found a more virtuous balance among the three sub-dynamics than in the more peripheral parts of the country. However, Figure 2 shows that the pattern "the farther from Moscow, the lower the synergy" has one striking exception, the Siberian District. Most probably, it is due to large university centers such as in Novosibirsk, Tomsk, and Krasnoyarsk. Figure 3 supports this explanation, but with Omsk instead of Tomsk as the local center of synergy.

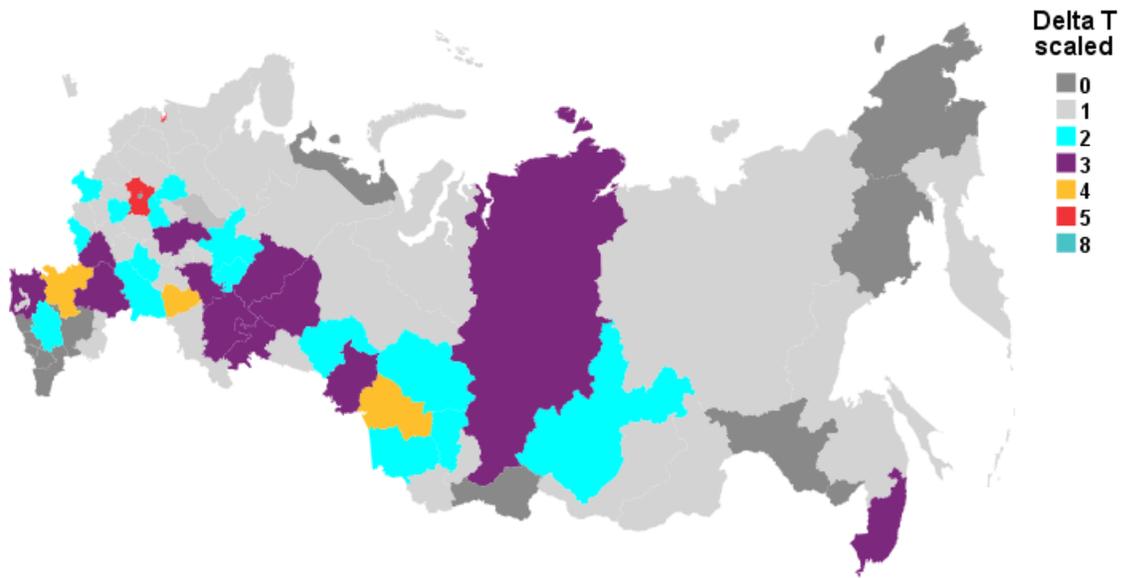

**Figure 3.** Synergies in the knowledge-based economy of Russia at the level of 83 Federal Subjects (2011; *N* = 594,987; mbits of information).

Scale:
- ∆T > 0;
- 0 > ∆T > -10 mbit;
- -10 > ∆T > -25 mbit;
- -25 > ∆T > -50 mbit;
- -50 > ∆T > -100 mbit;
- -100 > ∆T > -200 mbit;
- Moscow: - 484 mbit

Figure 3 provides a map of Russia with the 83 Federal Subjects (states) colored according to their respective contributions to the synergy in the knowledge-based economy. Figure 4 zooms in on the European part of Russia in Figure 3. This highlights the specific positions of Moscow and St. Petersburg. The scaling in the legend is similar as in Figure 3.

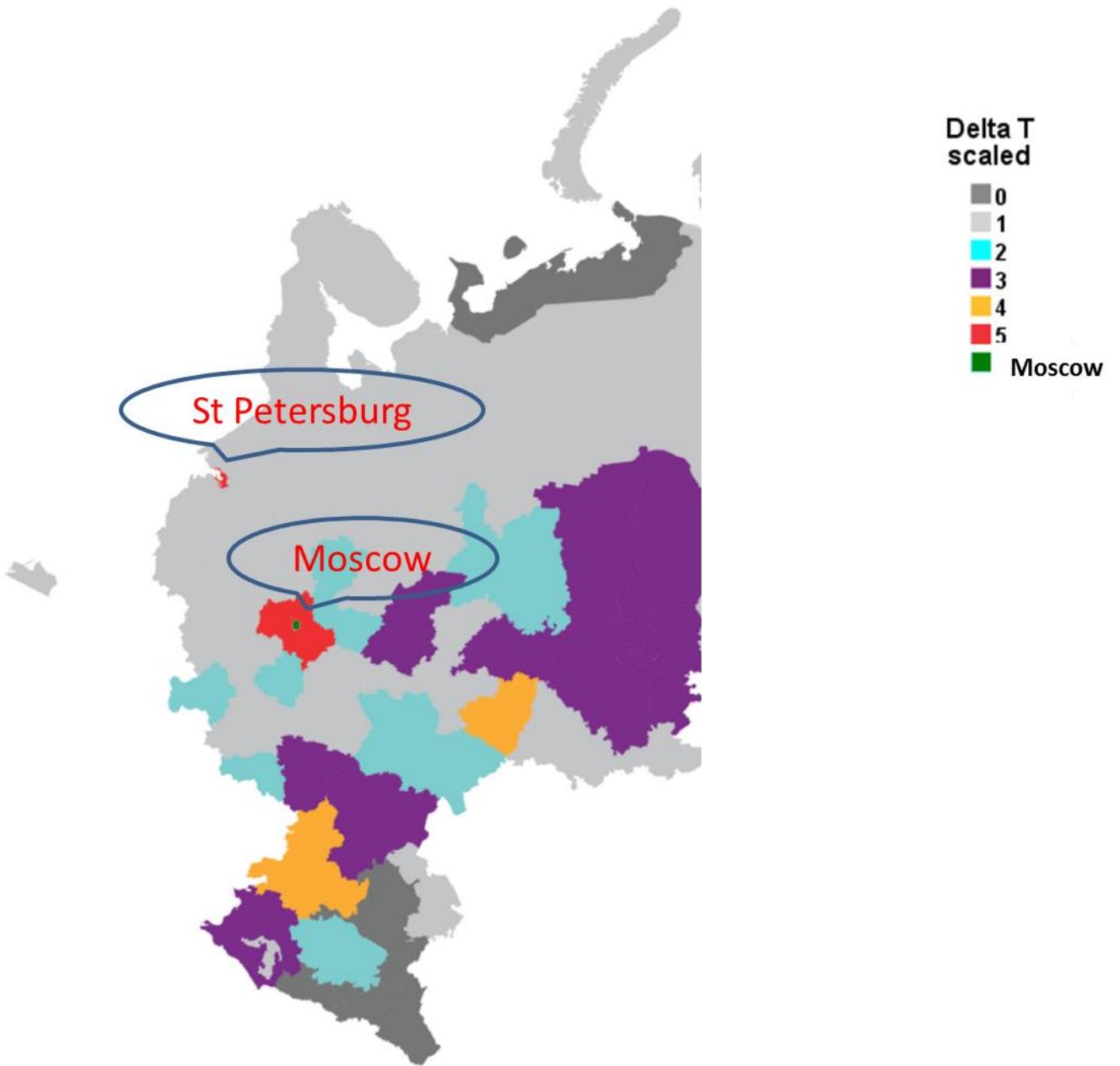

**Figure 4:** Synergies in the knowledge-based economy of the European part of Russia at the Federal Subject level (2011; *N* = 447,565 [75.3% of the 593,987 firms in the sample]).

The total synergy for the nation is –2690.7 mbits of information, of which 37.9% (–1019.7 mbits) is realized at the level above Federal Subjects (Table 7 column a: –704.2 mbits at the level of Federal Districts and –315.5 mbits at the national level). This 37.9% reduction of the uncertainty at the above-regional level is far more than the values reported for Norway (11.7%), the Netherlands (27.1%), or Sweden (20.4%).

**Table 7**: Synergy in Triple Helix interactions at different levels of administration; and in knowledge-based manufacturing (high- and medium-tech), and knowledge-intensive services (KIS); in mbits and as percentages.

|  | All sectors<br>N=593,987<br>(a) | High Tech<br>N=2,564 (0.4%)<br>(b) | Medium Tech<br>N=15,860 (2.7%)<br>(c) | KIS<br>N=76,078 (12.8%)<br>(d) |
|---|---|---|---|---|
| Federal Subjects (83) | -1670.9<br>(62.1%) | 879.3 | 555.6 | -1024.8<br>(54.9%) |
| Federal Districts (8) | -704.2<br>(26.2%) | -386.4 | -986.6 | -550.6<br>(29.5%) |
| National | -315.5<br>(11.8%) | -570.4 | -716.9 | -293.9<br>(15.7%) |
| Total | -2690.7 | -77.4 | -1147.3 | -1869.3 |
| normalized nationally | -2690.7 | -0.3<br>(0.01%) | -30.6<br>(1.1%) | -239.4<br>(8.9%) |

Table 7 concisely summarizes the results of this study decomposed both geographically and in terms of the most relevant sectorial division. First in column (a), the synergies generated at the levels of 83 Federal Subjects and eight Federal Districts, respectively, are summarized in terms of their respective contributions to the synergy at the national level. Considering firms in all sectors ($N = 593,987$), 62.1% of the synergy is found at the regional level of Federal Subjects; 26.2% at the level of the eight Districts; and 11.8% on top of that at the national level of the Russian Federation. Of these firms 19.9% were located in Moscow ($N = 118,072$); 33,489 (5.6%) in the surrounding Moscow Region; and 27,451 (4.6%) in St. Petersburg (Gini = 0.59). As was shown in Figure 4, these are by far the most integrated regions in terms of synergies between economics, technology, and geography.

In the second column (b) we focus on the 2,564 firms (0.4%) which are classified as "high-tech manufacturing" in terms of the classifications provided in Table 6. In this case, one third (33.3%; $N = 854$) of these firms is located in Moscow with another 7.0% in Moscow Region. St. Petersburg follows with 220 firms (8.6%). In the other regions, the presence of high-tech manufacturing is rare, that is, fewer than one hundred firms. In these sectors, the synergy is found at the national level more than at the district level. Only the Central Federal District (including Moscow) shows a synergy (-23.7 mbits; $N = 1,324$) at this next level of aggregation. At the regional level of Federal Subjects, the high-tech firms disturb the synergy considerably by generating 879.3 mbits of uncertainty in the regions. (We did not add percentages to this column because of the sign change involved.) In other words, high-tech manufacturing is not embedded at the regional level of Federal Subjects, but at the next level of Federal Districts (−386.4 mbits) and even more at the national level (−570.4 mbits).

Column (c) shows the results of focusing on the 15,860 firms (2.7%) that are classified as medium-tech manufacturing (in Table 6). In this case most of the synergy is found at the level of the eight Federal Districts, that is, above the regional level (−986.6 mbits); the national level adds another −716.9 mbits to this synergy. However, when normalized in relation to the national level, medium-tech manufacturing contributes only 1.1% (30.6 mbits) to the synergy in all sectors,[5] whereas they form 2.7% of the total number of firms.

---

[5] One can normalize the numbers at the national level by multiplication with $n_G/N$, or in this case 15,860 / 593,987.

In summary, medium-tech manufacturing does not add to the knowledge-base of the Russian economy, unlike other countries that were studied using this methodology hitherto. The distribution of medium-tech manufacturing across the regions is less concentrated than that of high-tech manufacturing. For example, 16.8% of them are in Moscow against 19.9% for the whole set (and 33.3% for high-tech manufacturing). In other words, medium-tech firms are distributed across the economy, but do not contribute to the synergy in the regional economies at the level of Federal Subjects.

Knowledge-intensive services (column d) are provided by 76,078 firms (12.8%) and contribute 8.9% to the synergy in the national economy. The coordination is both large at the level of the regions (54.9%), but heavily concentrated in Moscow as one of these regions ($N = 29{,}190$; $\Delta T = -738.8$ mbits, that is, 72.1% of the total of $-1024.8$ mbits). We conjecture that these services are provided in state-apparatuses and establishments related to these. They are pervasively present throughout the economy and integrating the economy. At the level of Federal Districts, only the North-Caucasian Federal District fails to show synergy at this level.

Table 8 lists the 15 Federal Subjects that contribute to the sample of KIS, in decreasing order of the (normalized) contributions to the synergy ($\Delta T$ in mbits). The values of $\Delta T$ seem to be strongly correlated (with the opposite sign) to the number of establishments ($N$), but for all 82 Subjects, the (Spearman) rank-order correlation between these two values is $-0.662$ ($p < .01$). In 52 of the 82 Federal Subjects, KIS has smaller numbers ($N < 1000$) and is not embedded regionally ($\Delta T > 0$).

| Federal Subject | N | ΔT in mbits |
|---|---|---|
| Moscow | 29109 | -738.84 |
| Saint Petersburg | 4445 | -88.58 |
| Moscow Region | 3755 | -55.71 |
| Samara Region | 1925 | -25.38 |
| Novosibirsk Region | 1842 | -22.09 |
| Sverdlovsk Region | 1787 | -21.61 |
| Rostov Region | 1478 | -17.48 |
| Perm Region | 1352 | -15.08 |
| Krasnoyarsk Region | 1218 | -10.88 |
| Republic of Tatarstan | 1205 | -9.75 |
| Omsk Region | 1124 | -8.58 |
| Udmurt Republic | 979 | -8.39 |
| Tomsk Region | 951 | -7.59 |
| Krasnodar Region | 1354 | -7.41 |
| Chelyabinsk Region | 1105 | -7.36 |
| …. | … | … |

**Table 8**: Fifteen Federal Subjects that contribute more than 1.2% to the number of KIS at the regional level; in decreasing order of the contributions to the synergy.

In summary, high-tech manufacturing plays a role exclusively in Moscow; medium-tech manufacturing is distributed across the country, but fails to contribute to the knowledge-based economy probably because these firms are not sufficiently embedded (Cohen & Levinthal, 1990). Where present in sufficient numbers, KIS is regionally integrated in Russia more than in other countries. On the basis of our previous studies, one would expect knowledge-intensive services and high-tech manufacturing to be more volatile ("footloose"; Vernon, 1979). We conjecture that knowledge-intensive services in Russia are related to the administration in the state apparatuses, and therefore not so flexible. One should also keep in mind that distances are large in Russia, making it difficult to offer services nation-wide or across regions.

**Conclusions**

The analysis of the Russian economy using the Triple-Helix indicator provides us with a perspective on an economy organized very differently from the Western economies that we have studied hitherto or the Chinese economy (Leydesdorff & Zhou, forthcoming). It transpires that the Russian economy is not knowledge-based. Synergies in the regions among existing technological and economic structures are disturbed instead of reinforced by medium-tech manufacturing and even more so by high-tech manufacturing. Knowledge-intensive services are grounded and not, as we hypothesized in the introduction (on the basis of previous studies), a mechanism that uncouples from the local economies. Both KIS and high-tech manufacturing are heavily centralized in Moscow.

Moscow, the Moscow Region, and St. Petersburg are the regions where synergy is generated to such an extent that the scale of the operations is different from those in the other regions or Federal Districts. The Federal Districts are a relevant level of coordination, but do not function as the provincial level in China to provide the main coordination mechanism of that economy (Leydesdorff & Zhou, in press). The centralization resembles that of Sweden (with Stockholm, Gothenburg, and Malmö/Lund providing 48.5% of the TH synergy; Leydesdorff & Strand, 2013). For Russia, however, the aggregate percentage for the top-3 regions (Moscow, Moscow Region, and St. Petersburg) is only 26.8% of the national synergy. In Sweden, the mechanisms of embedding high- and medium-tech manufacturing and KIS are similar to other European nations, whereas in Russia KIS is associated with governmental structures. High- and medium-tech manufacturing do not play a role in the coordination of the economy at the regional level.

It remains to be cautioned that our data is not based on official government statistics, but on a commercial database (*Orbis*™). *Orbis* data is collected by Bureau Van Dijk from numerous sources (more than hundred). These sources remain otherwise unrevealed to the user of this database (Ribeiro *et al.*, 2010). However, we are currently not aware of data of higher quality than this about the Russian economy in the three relevant dimensions and at the micro-level of establishments.

Given these *caveats*, one may wish to note in terms of policy implications that the Russian system of innovations integrates the three dimensions synergetically at all three levels of Federal Subjects (62.1%), Federal Districts (26.2%), and the nation state (11.8%; Table 7, column a). In other words, a Russian innovation system is in place at all three levels of administration. When the sectors are decomposed in terms of knowledge intensity, however, both medium- and high-tech firms are no longer integrated at the regional level of Federal Subjects. Medium-tech manufacturing contributes to the integration at the levels of districts more than nationally, whereas high-tech contributes mainly nationally (that is, in Moscow). Differently from Western Europe, knowledge-intensive services are embedded at all three levels (Table 7, column d), and function more or less comparably to the respective synergy for all sectors (Table 7, column a). Enhancing the circulation of these services and encouraging the diffusion of high-tech across the country—perhaps in the form of more competition—could be beneficial to the further development of a knowledge-based economy in Russia.

**Acknowledgements**
We are grateful to the referees for very rich comments. A previous version of this paper was presented at the 11$^{th}$ International Triple Helix Conference in London, July 6-8, 2013 (Perevodchikov *et al.*, 2013).